\documentstyle[12pt]{article}
\advance\hoffset by -1.1truecm
\advance\voffset by -1.0truecm

\newcommand{\x}{{\bf x}}
\newcommand{\p}{{\bf p}}

\begin{document}

\title{On Nonlocality, Lattices and Internal Symmetries}
\author{Achim Kempf\footnote{
Research Fellow of Corpus Christi College in the
University of Cambridge}\\
Department of Applied Mathematics \& Theoretical Physics\\
University of Cambridge\\ Cambridge CB3 9EW, U.K.}
\date{}
\maketitle
\begin{abstract}
We study functional analytic aspects of two types
 of correction terms to the Heisenberg algebra. 
One type is known to
induce a finite lower bound $\Delta x_0$ to the 
resolution of distances, a short distance cutoff
 which is motivated from string theory and quantum 
gravity. It implies the existence of families 
of self-adjoint extensions of the position operators 
with lattices of eigenvalues. These lattices, which
 form representations of 
certain unitary groups cannot be resolved on the
 given geometry. This leads us to conjecture that, 
within this framework, degrees of freedom that 
correspond to structure smaller than the 
resolvable (Planck) scale
turn into internal degrees of freedom with
 these unitary groups as symmetries. 
The second type of correction terms is 
related to the previous
essentially by "Wick rotation", and its basics are
here considered for the first time. 
In particular, we investigate unitarily 
inequivalent representations. 

\end{abstract}
\vskip-14.5truecm

\hskip10truecm
{\bf DAMTP-97-66} 

\hskip10truecm
{\bf hep-th/9706213}
\vskip14.5truecm
In the context of string theory and quantum gravity 
the possible existence of a
natural ultraviolet cutoff, e.g. at the Planck scale, 
has been widely discussed with various ansatzes, see e.g.
 \cite{townsend}-\cite{camelia}. In particular, these studies
include uncertainty relations of the form 
\begin{equation}
\Delta x \Delta p \ge \frac{\hbar}{2} (1+\beta (\Delta p)^2+...) \label{1}
\end{equation}
 with corresponding
corrections to the Heisenberg commutation relations of the form 
\begin{equation}
[\x,\p]=i\hbar(1+\beta \p^2+...) 
\label{ccr}
\end{equation}
Interest has so far mainly rested on the case 
 $\beta>0 $, as it is in this case that an 
ultraviolet regularising 
short distance behavior appears. In the generic case of corrections to $n$- dimensional commutation relations with 
Minkowski signature also correction terms of the type of $\beta<0$ 
are to be expected, at least for the temporal components. 
\medskip\newline
Let us therefore begin with a brief analysis of the case $\beta <0$.
The uncertainty relation 
yields $\Delta x_0 =0$, as usual.
However, taking the trace on both sides of Eq.\ref{ccr} shows
that finite dimensional representations are no longer excluded.
Indeed, there now exist even one-dimensional representations with $\x$
represented as some arbitrary number and $\p$ represented as
$\pm \vert \beta \vert^{-1/2}$. Indeed, 
all finite dimensional representations reduce to sums of 
these cases: In the $\p$ eigenbasis $\p_{ij}=\p_i\delta_{ij}$
 and the commutation relations, $\x_{rs}(\p_s-\p_r)
=i\hbar \delta_{rs}(1+\beta \p_r^2)$ yield $\p_r =
 \pm \vert \beta\vert^{-1/2}$, thus $([\x,\p])_{rs} = 0$, so that $\x$ is
diagonalisable simultaneously with $\p$, and we obtain $\p_{rs} 
= \mbox{diag}(p_1,p_2,...,p_n)$
and $\x_{rs} = \mbox{diag}(x_1,x_2,...,x_n)$ with $p_i \in 
\{-\vert \beta\vert^{-1/2}, \vert \beta\vert^{-1/2}\}$ and $x_i \in {I\!\! R}$. 
The infinite dimensional representations are harder to classify. Let us
 begin with the spectral representation of $\p$:
\begin{eqnarray}
\p.\psi(\lambda) &=& \lambda \psi(\lambda)\label{e1} \\
\x.\psi(\lambda) &=& i\hbar \left(\frac{d}{d\lambda}+\beta \lambda 
\frac{d}{d\lambda} \lambda\right)
\psi(\lambda)\label{e2}\\
\langle \psi_1\vert \psi_2 \rangle &=& \int_I d\lambda~ \psi_1^*(\lambda)
\psi_2(\lambda)\label{int}
\end{eqnarray}
We note that, as is easy to verify, exactly the family of operators $G$ defined through the
 integral kernel ($a,b \in C$)
\begin{equation}
G(\lambda,\lambda^\prime) =  \left( a \Theta(\lambda -
 \vert \beta\vert^{-1/2}) + b \Theta(\lambda + \vert \beta\vert^{-1/2})
\right) \delta(\lambda -\lambda^\prime) 
\end{equation}
commute with both $\x$ and $\p$. Each $G$  is diagonal and constant apart from two steps
where it cuts momentum space, and with it the representation, into three unitarily inequivalent parts.
The representation which has the proper limit as $\beta
\rightarrow 0$ is given by Eqs.\ref{e1}-\ref{int} 
with the integration interval
$I:= I_c = [-\vert \beta\vert^{-1/2}, \vert \beta\vert^{-1/2}]$. Thus, $\p$ becomes a bounded self-adjoint operator. Let us  
calculate the defect indices of $\x$ in this representation, i.e. 
the dimensions of the kernels of $(\x^*\pm i)$, i.e. we
check for square integrable solutions to (from now on we 
set $\hbar =1$)
\begin{equation}
i \left(\partial_\lambda -\vert \beta\vert(\lambda^2 \partial_\lambda
 +\lambda)\right) \psi_\xi(\lambda) = \xi \psi_\xi (\lambda)
\end{equation}
with $\xi = \pm i$. The equation is solved by 
\begin{equation}
\langle \lambda \vert \xi\rangle =
\psi_\xi(\lambda) = N \left(1-\vert \beta\vert \lambda^2\right)^{-1/2} 
\left(1- \sqrt{\vert \beta\vert} \lambda\right)^{i \xi/2} 
\left(1+\sqrt{\vert \beta\vert} \lambda\right)^{-i \xi/2} 
\label{sols}
\end{equation}
which are non-square integrable on $I_c$ for all $\xi \in \vert\!\!\!C$, in particular also for $\xi=\pm i$. Thus, 
the defect indices are $(0,0)$, i.e. $\x$ is still essentially 
self-adjoint with a unique spectral representation (recall that 
the operator $i\partial_\lambda$ which normally represents $\x$ on momentum space has defect indices (1,1) on the intervall).
The position eigenfunctions
are given by Eq.\ref{sols} for real $\xi$.
With the continuum normalisation
  $N=(2\pi)^{-1/2}$ it is not difficult to verify orthonormalisation and completeness:
\begin{eqnarray}
\int_{-\vert\beta\vert^{-1/2}}^{\vert\beta\vert^{-1/2}}
d\lambda ~ \langle \xi \vert\lambda\rangle\langle\lambda\vert\xi^\prime
\rangle &=& \delta(\xi-\xi^{\prime})\\
\int_{-\infty}^\infty d\xi~ \langle 
\lambda\vert\xi\rangle\langle\xi\vert\lambda^\prime\rangle &=& 
\delta(\lambda - \lambda^\prime)
\end{eqnarray}
The generalised Fourier factor given in Eq.\ref{sols} yields the  transformation
 $\psi(\xi) =  \int_{I_c} d\lambda
\langle \xi\vert\lambda\rangle \psi(\lambda)$
that maps momentum space wave functions $\psi(\lambda)=\langle \lambda \vert \psi\rangle$ 
to position space wave functions $\psi(\xi) = \langle \xi \vert \psi \rangle$. To summarise, we have found
 no short distance cutoff in positions, while we have found that
momentum space becomes bounded.
\medskip\newline
Let us now turn to the case $\beta>0$.
As is well known, and as is easily derived from Eq.\ref{1},
the position resolution $\Delta x$ now becomes finitely bounded from below: $\Delta x_0=\hbar \sqrt{\beta}$. To be precise,
for all normalised vectors $\vert \psi\rangle $
in a domain $D$ on which the 
commutation relations are represented the
position uncertainty obeys $\Delta x_{\vert\psi\rangle} =\langle \psi \vert (\x -\langle\psi\vert \x
\vert\psi\rangle)^2\vert\psi\rangle^{1/2} \ge 
\Delta x_0 $. A convenient representation is given by 
Eqs.\ref{e1}-\ref{int} with $I=$R.
Technically, on any dense domain $D$ in a Hilbert space $H$
 on which the commutation relations hold the position operator can only be symmetric
 but not self-adjoint, as diagonalisability is excluded by the
 uncertainty relation (eigenvectors to an observable  automatically have
vanishing uncertainty in this observable). This also excludes the possibility of finite 
dimensional representations of the
 commutation relations (since in these symmetry and self-adjointness coincide), 
as could of course also be seen by taking the trace of both sides of Eq.\ref{ccr}. Generally, 
in order to insure that expectation values of positions and momenta are real, we only consider corrections to the commutation relations which are consistent with an involution which
acts on the generators as $\x_i^*=\x_i, \p_i^*=\x_i$. 
The involution then also insures that the 
deficiency indices of the $\x_i$ (and $\p_i$) on any dense domain $D$ on which the
 commutation relations hold are equal, implying that the $\x_i$ do
have self-adjoint extensions in $H$, though not in $D$. This 
functional analytic structure was first found in
\cite{ixtapa,ak-jmp-ucr}. 
The self-adjoint extensions now have been
calculated explicitly for a number of cases, for our one-dimensional case here, in \cite{kmm}. 

We remark that a finite lower bound on the standard deviation in positions
is, interpretationally, an ensemble-based 
short distance regularisation (which could only appear in
quantum theory). There exists a straightforward way of introducing these generalised commutation relations
into the quantum field theoretical path integral \cite{ft} with 
the functional analysis of representations on wave functions extending to representations on fields (though the interpretation does of course not extend straightforwardly). We will in the
following use the quantum mechanical rather than the field
theoretical terminology, the analysis is the same.
It has been shown that 
this ensemble cutoff does indeed regularise the ultraviolet 
in euclidean field theory \cite{ft}-\cite{km}.
Let us now discuss further physical 
implications, related to internal symmetries. 
\medskip\newline
As we will see, 
the unobservability of localisation beyond 
the minimal uncertainty $\Delta x_0$ can be seen to represent
a local symmetry, where degrees of freedom which correspond to small scale structure beyond the
Planck scale turn into internal degrees of freedom.

Consider a 
$*$-representation (such as given by Eqs.\ref{e1}-\ref{int}) of the commutation relation Eq.$\ref{ccr}$ 
on a maximal dense domain $D$ in a Hilbert space $H$. 
Then $\x$ is merely symmetric, i.e. $D$ is smaller than the domain $D_{\x^*}$ of the 
adjoint operator $\x^*$ (which is not symmetric). The deficiency 
spaces $L_+,L_-$, i.e. the spaces
spanned by eigenvectors of $\x^*$ with eigenvalues
 $+i$ and $-i$ are one-dimensional, i.e. the deficiency indices are (1,1) (also in $n$ dimensions they are equal,
 due to the involution).

Thus, there exists a set of 
self-adjoint extensions of $\x$ which is in
one-to-one correspondence with the set of unitary transformations $\tilde{U} : L_+ \rightarrow L_-$. (Recall that, by the usual procedure, each $\tilde{U}$
defines a unitary extension of the Cayley transform of $\x$, with the inverse Cayley transform then defining 
a self adjoint extension of $\x$. On the eigenvalues, Cayley transforms are Moebius transforms.)

 The $\tilde{U}$ differ exactly by
the set $G$ of unitary transformations $U: L^+ \rightarrow L^+$.
In general, for defect indices $(n,n)$,
this is the unitary group $U(n)$,  
which we may here call the local group $G$. Thus, the set of self-adjoint extensions $\{\x_\alpha\}$ 
forms a representation of the local group. $\alpha$, which labels the
self adjoint extensions, is a
vector in the fundamental representation of $G$.
The local group also acts on the
 set of spectra $\{\sigma_\alpha\}$
of the $\x_\alpha$. Let us denote the eigenvalues of the self-adjoint
extension $\x_\alpha$ by $v_\alpha(r)$. Then, for any fixed $r$,
we obtain an orbit $O(r):= \{ v_{U.\alpha}\vert U\in G\}$
 of eigenvalues under the action of $G$. 

Let us consider the example of the one-dimensional
case above. The scalar product of eigenvectors of $\x^*$ has been
calculated in \cite{kmm}:
\begin{eqnarray}
\langle \xi \vert \xi^\prime \rangle  = 
 \frac{2 \sqrt{\beta}}{
\pi (\xi - \xi^{\prime})} {\mbox{ }} \sin\left(\frac{\xi 
-\xi^{\prime}}{2 \sqrt{\beta}} 
\pi\right)
\label{spfp}
\end{eqnarray}
From its zeros
 we can read off the family of discrete spectra of the self-adjoint extensions:
\begin{equation}
\sigma_\alpha = \left\{v_{\alpha}(r) = 
(2 r + s/\pi) \sqrt{\beta}~\vert~
r \in \mbox{N}\right\} ~~~~~\mbox{where}~~~~~
\alpha = e^{is} \mbox{~~ with ~~}s \in [0,2\pi[
\end{equation} 
The spectra are equidistant and two self-adjoint extensions
only differ by a shift of their lattice of eigenvalues. The local group is here
the group of translations of the lattices of eigenvalues. 
Due to the periodicity of the lattice this group is topologically
$S^1$, or $U(1)$. This reflects that in this case
$L_+$ is one-dimensional and the
 self-adjoint extensions therefore form a representation of the
local group U(1). 

Each choice of self-adjoint extension of the position operators therefore
corresponds to a choice of lattice on which the physics takes place. However, the commutation relations also
imply that the smallest uncertainty in positions becomes finite and large enough so that the actual choice
of lattice cannot be resolved. Technically, all
self adjoint extensions of $x$ coincide when restricted to a domain $D$ on which the commutation relations hold.

If, therefore,
 with a physical state $\vert \psi\rangle \in D$ also some vector $\alpha$ is specified, as a
choice of self-adjoint extension, the action should be invariant,
i.e. we arrive at a global symmetry principle.
The additional information given by $\alpha$ can be interpreted. Assume that the state of a particle is projected onto 
a state of maximal localisation ($\Delta x = \Delta x_0$)
with position expectation $\xi$. Specifying $\alpha$ is to specify one point in the orbit of
the eigenvalue $\xi$ under the action of the local group. As a convention one can specify that this is  
where the maximally localised particle is said to "actually" sit. This is consistent because    
the radius of the orbits of the eigenvalues 
is $\sqrt{\beta} = \Delta x_0$ i.e. of the size of the finite minimal uncertainty $\Delta x_0$, so that all these conventions,
differing only by the action of the local group, cannot be  distinguished observationally. 
For example, the pointwise multiplication of fields as discussed
e.g. in \cite{kmm} can be reformulated in terms of a choice of position eigenbasis, 
rather than the set of maximally localised fields. The gauge principle is that the action is invariant under
the local group. We remark that the proof of ultraviolet 
regularity will still go through, since not only the
fields of maximal localisation, but also the position eigenfields are normalisable. 

On the other hand, the `local' group may also be taken to act
locally, i.e. we consider $\vert \psi\rangle \in D \otimes L_+$.
It is unobservable whether one specifies one self-adjoint extension's lattice here and another's
there, as long as the parallel transport of $\alpha$ is consistently defined. At large scales this should lead
to the ordinary local gauge principle. 
We note that the the local gauge group will be determined
through the functional analysis of the position operators. This 
in turn depends on the choice of short distance structure as specified through the 
corrections to uncertainty- and commutation relations. 
In the physical case of the Minkowski signature
further nontrivial structures can be expected
to arise from the behavior of the coordinates which behave according to the 
$\beta <0$ case, as discussed above. 

To summarise,  there is a possibility that internal
 symmetry spaces arise as 
deficiency spaces of position operators.
 Introducing $\Delta x_0>0$ 
the infinite dimensional Hilbert space of fields develops
 special dimensions that
correspond to degrees of freedom that describe localisation
beyond what can be resolved, and which can therefore be 
viewed as internal degrees of freedom. 
Basically, the idea is, that certain corrections to the 
uncertainty relations lead to physics on a whole set of possible
lattices, while the choice of any particular lattice from the set
cannot be resolved and does therefore correspond
 to an internal degree of freedom. This may be a new mechanism, or it may be 
a reformulation of the Kaluza Klein
idea, in which case one may expect a deeper relation to string theory.


\begin{thebibliography}{99}
\bibitem{townsend} P.K. Townsend, Phys. Rev. {\bf D15}, 2795 (1976)
\bibitem{grossmende} D.J. Gross, 
P.F. Mende, Nucl. Phys. {\bf B303}, 407 (1988)
\bibitem{sm-cqg} S. Majid, Class. Quantum Grav. {\bf 5}, 1587 (1988)
\bibitem{amati} D. Amati, M. Ciafaloni, G. Veneziano, Phys. Lett. 
{\bf B 216}, 41 (1989)
\bibitem{konishi} K. Konishi, G. Paffuti, P. Provero, Phys. Lett. 
{\bf B234}, 276 (1990) 
\bibitem{ixtapa} A. Kempf, Proc. XXII DGM Conf. Sept.93 Ixtapa
(Mexico), Adv. Appl. Cliff. Alg (Proc. Suppl.) {\bf (S1)} (1994)
\bibitem{maggiore} M. Maggiore, Phys. Lett. {\bf B319}, 83 (1993)
\bibitem{ak-jmp-ucr} A. Kempf, J. Math. Phys. {\bf 35} (9), 4483 (1994)
\bibitem{jaeckel} M.-J. Jaeckel, 
S. Reynaud, Phys. Lett. {\bf A185}, 143 (1994)
\bibitem{ahlu} D.V. Ahluwalia, Phys. Lett. {\bf B339}, 301 (1994)
\bibitem{garay} L.J. Garay, Int. J. Mod. Phys. {\bf A10}, 145 (1995)
\bibitem{kmm} A. Kempf, G. Mangano, 
R.B. Mann, Phys. Rev. {\bf D52}, 1108 (1995), hep-th/9412167
\bibitem{doplicher} S. Doplicher, K. Fredenhagen, J.E. Roberts,
Comm.Math.Phys. {\bf 172}, 187 (1995)
\bibitem{witten} E. Witten, Phys. Today {\bf 49} (4), 24 (1996)
\bibitem{luk} J. Lukierski, Preprint hep-th/9610230
\bibitem{camelia} G. Amelino-Camelia, gr-qc/9706007
\bibitem{ft} A. Kempf, Preprint DAMTP/94-33, 
hep-th/9405067, and Czech. J. Phys. (Proc. Suppl.), {\bf 44}, 1041 (1994)
\bibitem{ak-jmp-reg} A. Kempf, Preprint hep-th/9602085,
J. Math. Phys. 38, 1347 (1997)
\bibitem{ak-prd-reg} A. Kempf, Phys. Rev. {\bf D54}, 5174 (1996),
hep-th/9602119
\bibitem{go} A. Kempf, presented at 21st International
Colloquium on Group Theoretical Methods in Physics 
(ICGTMP 96), Goslar, Germany, July 1996, hep-th/9612082, DAMTP-96-101
\bibitem{km} A. Kempf, G. Mangano, Phys. Rev. {\bf D55},
 7909 (1997)
\end{thebibliography}
\end{document}